\documentclass[12pt]{article}
\usepackage{epsfig}

\usepackage{amsmath}
\usepackage{amssymb}
\usepackage{euscript}

\begin{document}

% title
%%%%%%%%%%%%%%%%%%%%%%%%%%%%%%%%%%%%%%%%%%%%%%%%%%%%%%%%%%%%%%%%%%%%%%%%%%%%%%%
%%%%%%%%%%%%%%%%%%%%%%%%%%%%%%%%%%%%%%%%%%%%%%%%%%%%%%%%%%%%%%%%%%%%%%%%%%%%%%%
\begin{titlepage}

 \vspace{10mm}

\begin{center}
{\LARGE\bf A Gauge Model 
\vspace{3mm}
of Data Selection, Acquisition 
\vspace{3mm}
and Analysis for LHC$^{\star}$}

\end{center}

\vspace{15mm}

%\vspace*{6mm}
%\vspace*{6mm}
%\vspace*{6mm}
% authors

\begin{center}
{\large\bf  Mieczyslaw Witold Krasny } \\

\vspace{5mm}

LPNHE, Pierre and  Marie Curie University, Paris \\
and  \\
Institute of Nuclear Physics,  HNINP-PAS, Cracow

\end{center}

\vspace{25mm}
\begin{abstract} 
A novel model of the data selection, acquisition and analysis for a multi-purpose and multi-component
high-energy-physics  experiment is presented.
Its departure point is the freedom and the responsibility given to the different physics groups
of the experiment  to impose, on the {\it event-by-event basis},  their  
physics-goal-optimal configurations of (i) the sub-detectors, (ii) the trigger and data acquisition system, and  
(iii) the reconstruction and analysis framework. Its target is to develop, in a close  analogy to 
the construction of the gauge models in particle physics, the overall data handling scheme, in which a multi-purpose experiment becomes an association of coexistent, yet largely independent, physics-group-based sub-experiments sharing common hardware maintenance, data-acquisition,
and data reconstruction resources.  
\end{abstract}

\vspace{20mm}
\footnoterule
\noindent
{\footnotesize
$^{\star}$ Contribution to  the Cracow Epiphany Conference on LHC Physics, 4-6 January 2008, Cracow, Poland
}

\end{titlepage}

\section{Introduction}

In the pre-LHC  multi-purpose high-energy-physics collider experiments,
the process of diversification of the physics-goal-dedicated data analysis methods has been largely decoupled
from the process of the data taking. The experiment's physics-groups could develop  their optimal selection criteria for the recorded data. However, these criteria had to be confronted with, often mutually exclusive, requirements of the other physics groups and compromises had to be found while constructing global trigger menus and the 
data selection algorithms. Moreover,
within such a scheme  the raw data content of recorded events and  the framework of 
their (i) initial on-line selection, (ii) reconstruction, and (iii) analysis were the same for all the events. The above scheme  assured an``offline-analysis-friendly" encapsulation of the technical aspects of the data taking process
and has been generally considered as optimal because 
it has never been over-restrictive for the physics-group-optimal data selection and data analysis methods.  

The initial configuration of the multi-purpose LHC experiments is based on a continuous 
extrapolation of the above paradigms to the LHC experimental environment. One of the  
basic questions which is worth addressing in the advent of the LHC experimental program is: 
"Will such a paradigm survive the``LHC phase transition"  to  a new detector-operation regime, characterized by a small, $O(10^{-7})$, ratio of  the rate of the recorded events to the 
rate of the proton-proton collision events, and to a new sociological regime of a large,  $O(10^3)$, group of  physicists  exploring the new energy frontier using the diverse analysis methods?". 

The starting   point of our proposed  ``Gauge"  Model\footnote{The terminology ``Gauge Model" 
is to draw attention to the analogy with the well-known gauge models of particle interactions, where,
often, we may exploit the gauge freedom by making a specific choice of gauge that greatly simplifies the 
analysis of a particular problem.} 
 of Data  Selection, 
Acquisition and Analysis  is the observation   
that the above paradigm  cannot be 
continuously extended to the LHC environment,  without
significant sacrifices in the scope and in the quality of its 
experimental program. Physics groups analyzing rare  events would clearly prefer  to have an access to the 
full raw electronic data. Such data cannot be recorded for 
{\it all the events}  because of the  bandwidth limits of  the data transmission. 
Physics groups attempting to select events based on exclusive topological criteria would prefer
to organize the High Level Trigger (HLT)  algorithms within a framework which is clearly distinctive to the optimal one for  selection of events  on the basis of the  inclusive trigger objects.
The optimal  trigger-frequency-versus-event-length trade-offs,  for both the detector monitoring events and 
for the physics events,  are often contradictory for groups analyzing large cross-section processes  and those 
searching for rare phenomena.   
Physics groups searching for new physics would certainly prefer to optimize the event selection capacity 
at the cost of the measurement precision. Physics groups working on high precision measurements would
clearly opt for an opposite solution. 

The proposed model tries to inject  a sufficient flexibility to  the data taking and data analysis 
process such that the optimal choices  could be made independently by each of the physics groups.
At the heart of the proposed model is the freedom and responsibility, given to the physics groups, 
to effectively impose the data-taking configuration of the sub-detectors and of the 
Trigged and the Data Acquisition (TDAQ)  system
{\it on event-by event basis}  allowing for the  physics-group-optimal use
of the detector capacities. As an example,
a physics group analyzing rare, large $E_T$  events 
is given the freedom to impose 
registering the most complete front-end electronic information 
of each of the  sub-detectors. This  group may implement   
a strategy based on inclusive, low purity but  high efficiency, 
region-of-interest-guided on-line selection methods.  {\it Within the same run},  
a physics group interested in large cross section
processes is given a freedom to impose
registering   highly-compressed front-end electronic information
restricted to  a subset of sub-detector partitions. This group 
may employ  the on-line event selection methods based upon 
global topological
criteria,  rather than based upon the region-of-interest  guided inclusive selection criteria, 
and may optimize the purity of the selected sample, rather than 
the efficiency of the event  selection. A flexibility  is given 
to each of the groups to run, if necessary, the group-optimal  
software selection-framework  implemented on a subset 
of the the LVL2 and  EF processors, and the group-optimal 
offline data analysis and data access framework implemented 
on a subset of the Tier 1 and Tier 2 processors.  

The proposed model may be applicable to the advanced  phase of the LHC experimental program, 
if  the quest for the best specific-physics-program-oriented
use of the detector and of the LHC capacities -- 
confronted with the hardware, software,  and sociological  complexity
of the LHC experiments -- ends-up in a phase-transition in the data handling paradigms.
Following such a phase transition,  the LHC experiments could metamorphose,  
becoming  associations of coexistent  yet largely independent  (transparent)
physics-group-based  sub-experiments,   sharing 
common hardware  maintenance, data acquisition, calibration,
and reconstruction resources. 
For the  author of this note,  such a phase transition is inevitable --
if the full discovery potential of LHC is to be exploited.
It may be accelerated if 
the initial discovery scenarios fail and/or  if there 
will  be a sufficient pressure of the physics-groups to implement  their group-specialized methods of 
the data selection and analysis. 

The proposed model is constructed according to  the construction pattern
of the gauge models in particle physics.
In  these  models the``gauge-choice" is arbitrary. It could reflect: 
the technical simplification of the calculations, a wish to preserve the  physics 
interpretation of the intermediate calculation steps,  or simply,  a specific physicist- taste.
The only constraint  is that the final results of the model calculations are gauge independent. In a close analogy,  
the proposed model allows for a ``gauge-dependent" 
event-by-event  selection of: the format and the content of the detector raw data,  and of the event building 
method. Moreover, it  gives a freedom 
of:  a ``gauge-dependent",  event-by-event choice of the event-selection framework
 implemented on one of the slices of  the TDAQ system,  and of a ``gauge-dependent", 
event-by-event choice of the offline data analysis framework implemented
on one of the slices of  the offline computing grid.   The full chain of data selection and analysis
becomes thus  ``gauge-dependent" while the  physics results remain to be ``gauge-invariant". 
In addition,  the proposed model inherits from the gauge models the handling 
methods of the gauge symmetry breaking phenomena reflecting  the detector-hardware  
operation constraints. 

Inevitably, the  model presented in this note changes the invariant 
character of the content of the registered raw data and of the event selection efficiencies from the 
experiment-invariant  one to the physics-group-dependent one.
The notion of a ``experiment-invariant"  data is thus 
lost and replaced by the notion 
of``physics-goal-optimal" data.
The recorded data preserve, however,  the experiment-wide universality
of the event structure, which is mandatory  for preserving the universality of the 
common, low-level,  event reconstruction,  detector calibration 
and the data quality monitoring software.
  
\begin{figure}
\begin{center}
\leavevmode
\epsfig{file=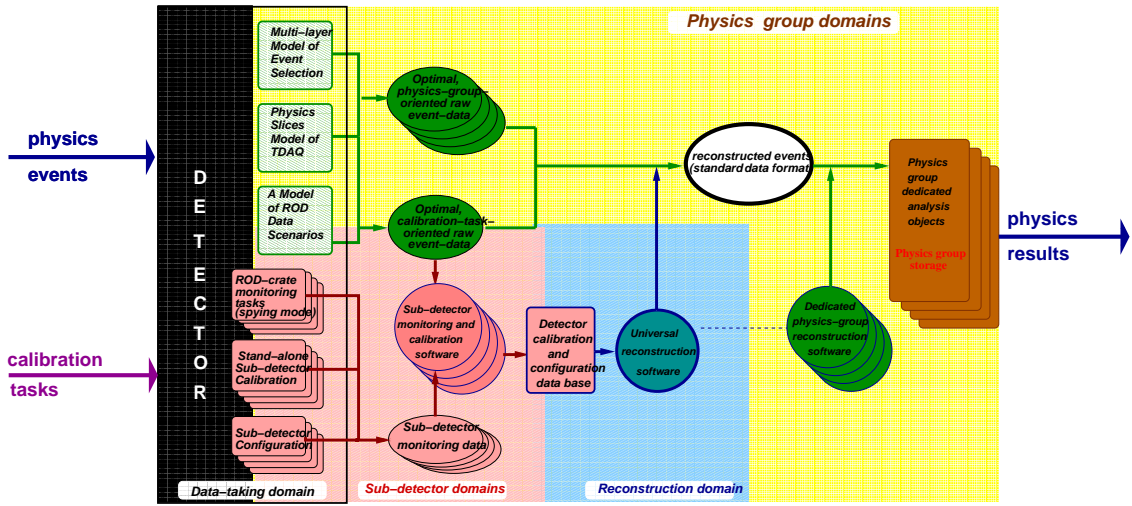,width=15cm}
\vspace{1.0cm}
\end{center}
\caption{The responsibility domains.}  
\label{data}
\end{figure}

The model changes significantly the borders and the interfaces between
the domains of the central data acquisition, sub-detectors, general data 
reconstruction and physics groups. Each of the physics groups 
can choose independently its preferred raw data form, the event  selection 
and event building strategy, the hardware capacity of their TDAQ slice, and
its  optimal data reconstruction software on a grid-slice in all the aspects except 
for the detector calibration and the low-level event reconstruction. 
The disaggregation of the data taking and data analysis process
into the domains of responsibilities is shown in Fig.\ref{data}.

This contribution to the Epiphany Conference presents the basic functionalities 
of the model as  constructed in the years 2002-2003 and as projected to the
hardware capacities of the ATLAS experiment   \cite{ATLAS}. 
Its details were documented in a series of notes \cite{TDAQnotes}. 
Introducing novel  ideas, in the year 2003, i.e. in  the phase of the
implementation of the present, widely accepted,  data handling 
paradigms into their concrete realizations was not, perhaps quite rightly,  considered as 
a constructive  action. This may no longer remain true in the forthcoming data taking phase of the 
LHC physics program when both the merits and the bottlenecks of the presently implemented 
data handling methods will become apparent. Following such a  phase
the LHC data taking and data analysis paradigms may need to be overhauled and, 
in such a case,  the proposed model could serve as an example of alternative solutions.

\section{ The Adaptive Variability Domains}  
 
The proposed model assumes the 
following definition of the function 
of the LHC  detectors:
{\bf The function of the  LHC detectors  consists of the selection of 
those of collision events 
which  enlarge our understanding of particle interactions,  and
of recording only  those of the  event data which  
are  sufficient for the most statistically and systematically 
precise measurements.}

\begin{figure}
\begin{center}
\leavevmode
\epsfig{file=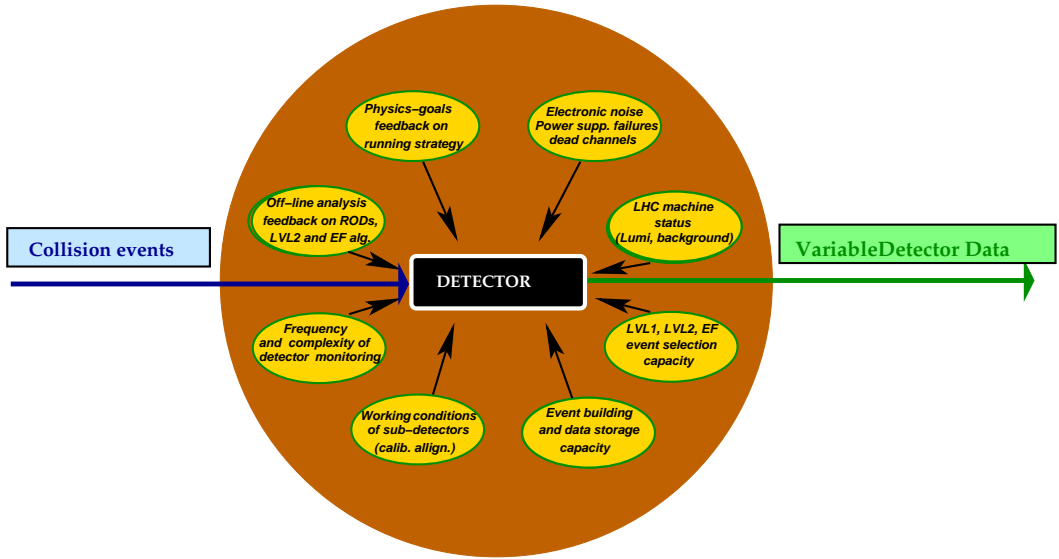,width=14cm}
\vspace{1.0cm}
\end{center}
\caption{The detector operation environment.}  
\label{environment}
\end{figure}

In order to fulfill this goal,   
all  the experiment sub-components (the sub-detectors, the TDAQ, the software domain,
the detector-performance-control  domain,     
and the physics analysis domain)   have to develop their specific adaptive capacities
to best cope with the variable data-taking environment illustrated in Fig. \ref{environment}.
Each sub-component could, in principle, develop individually its
adaptive functions. 
The resulting {\it organism} would however
be  {\it biologically primitive} and thus {\it evolutionary unstable}.

One of the goals of the proposed model  is to
document an attempt to define  the adaptive 
capacities of the experiment  
in a close analogy to  
advanced biological systems. In such  systems 
the function of adaptation to external environment
is delegated to specialized organs.
Handling of most of environmental changes can be then largely 
confined to these organs. 
This allows for
reducing the reaction time 
to external environmental stimuli. 

Such an adaptation form 
is characterized by  an  encapsulation 
of the {\it reflex-type} functions of the organism.
As a consequence, it leads to  reduction of the 
reaction time to those of the environmental  stimuli
for which the coherent response of the whole organism is indispensable.
Such specialized ``organs'' will be 
called in this note {\bf variability domains}.

The seed and central point  of the proposed model is the
conjecture that 
the adaptive capacity of the data-taking and on-line data analysis
process to  the environmental constraints
can be delegated  to the following three  variability domains: 
\begin{enumerate}

\item 
 the Read-Out-Driver (ROD) data variability domain; 
\item
the event selection tools variability domain;
\item
the TDAQ-slices  configuration variability domain. 

\end{enumerate}

These three variability domains are modeled in the following 
sections of this note. 

\section{The Model}

\subsection{The construction steps}

\begin{figure}
\begin{center}
\leavevmode
\epsfig{file=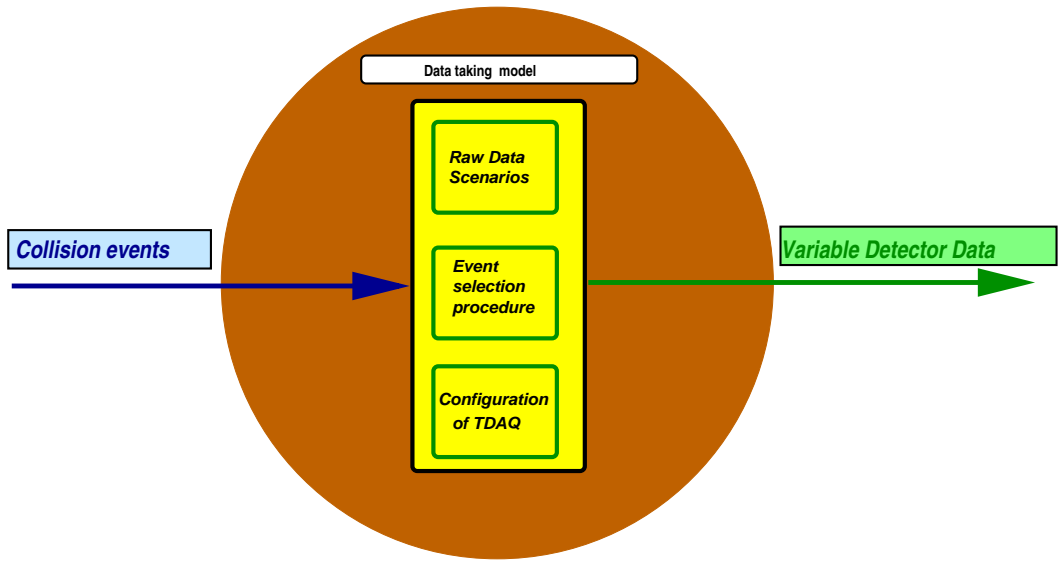,width=14cm}
\vspace{1.0cm}
\end{center}
\caption{The data-taking model.}  
\label{model}
\end{figure}

The model of data-taking is illustrated in Fig. \ref{model}.

Any offline task, attempting to analyze  collision 
events,  
instead of being confronted with sophisticated 
and variable detector operation environment depicted in Fig. \ref{environment}, 
will be exposed only to precisely-defined 
quantum-reactions  of the data-taking 
process to environmental changes.
These reactions will be fully encapsulated 
within the three variability domains and 
will be represented by quantum transitions 
between the allowed discrete set of states.
The model is constructed in three steps:
\begin{enumerate}
\item
The fist step consists of 
defining of the complete set of  {\it eigenstates}
of each of the three variability domains.
\item
The second step consists of projecting  these eigenstates 
onto the detector-partitions and the TDAQ-partitions granularity.
\item
The third step consists of  specifying the model dynamics  
in terms of  the causality and the time-granularity  pattern of quantum transitions between 
the allowed eigenstates.
\end{enumerate} 

The above three steps are described in the following three sections.

\subsection{The Eigenstates}

\subsubsection{The  eigenstates  of the ROD data} 

The eigenstates of the variability domain of the ROD data 
are specified in terms of: 

%The  ROD data variability domain is modeled   
% in terms of allowed changes in : 
\begin{itemize}
\item
the  data content;
\item
the data compression  method;
\item 
the  zero suppression scheme; 
\item
the channel addressing mode; 
\item
the format and the content of the ROD summary blocks. 
\end{itemize}

Let me give few concrete examples of the ROD data eigenstates 
for the ATLAS experiment \cite{ATLAS}.

The full bit history of the Transition Radiation Tracker (TRT)-straw 
signals passing 
low and high thresholds, and 
the leading-edges bit-history are he examples 
of the TRT data-content eigenstates.

Packing Tile Calorimeter (Tile-Cal) cell energies  
in 11 bits,  packing of
adjacent Semiconductor Tracker (SCT) strip info  into 32 bit words
are examples of the data compression eigenstates.\footnote {Data compression
does not change the information encoded in  the raw data 
but may result in variable unpacking methods 
of the byte-streams of the raw detector data.}

Zero suppression scheme consists of  
dropping fully  or partially  
the detector channel info which is unlikely
to be correlated with the  passage of a particle 
produced in the beam-beam  collisions. For example, dropping invalid  TRT straws,
dropping the $2 \sigma$  ``electronic noise'' Liquid Argon Calorimeter (LAr) channels,  or skipping 
the info which may be derived from the neighbor
channels ( e.g.,  the strip hits within a continuous cluster) are the 
possible eigenstates of the zero suppression scheme.
%\newpage
%width=10cm,bbllx=-29988 ,bblly=274, bburx=1012 ,bbury=30780}
%bbllx=-29988 ,bblly=274, bburx=1012 ,bbury=30780}

The sub-detector-optimal eigenstates of the addressing modes 
include  
direct addressing  of non-suppressed
channels (silicon trackers) or   bit pattern 
addressing mode of the LAr and Tile-Cal cells.
The addressing modes are correlated 
with the data compression modes  and with 
the zero suppression schemes.

The concept of the ROD summary blocks is new.
The ROD summary blocks contain the LVL2 trigger dedicated
information which will be used by the  ultra-fast  selection
algorithms. 
LAr fixed-format  blocks
containing the bit pattern of the energy-threshold crossing 
is an example of the LAr summary block eigenstate.

A central  point of  
 the proposed model is that 
a broad spectrum of sub-detector-allowed operation modes 
is confined to a {\bf small}  set of eigenstates 
spanning the full space of compromise between compactness and  completeness
of information. 
 These eigenstates
are referred  to as {\it data scenarios}.
Reactions of the sub-detector operation modes  to variable data  taking 
environment  are {\bf confined}  in the proposed model, to 
transitions (quantum-jumps)  between the allowed data scenarios.

\subsubsection{The eigenstates of the event-selection tools}

The eigenstates of the variability domain of the  event selection tools are specified in terms of: 

\begin{itemize}
\item 
the allowed Level 1 (LVL1), Level 2 (LVL2) and Event Filter (EF)  trigger signatures; 
%trigger elements(including global veto) threshold, prescalefactors
\item
the allowed layers of the event-selection algorithms.  
\end{itemize}

The trigger signatures  are modeled in the standard  way  \cite{TDR_TDAQ}
in terms of trigger elements, trigger thresholds,
and pre-scale factors.  
   
In order to cope with the dynamic  data-taking  environment  
new species  of event selection algorithms are proposed.
They extend the list of  algorithms presented in \cite{Steve},
and based solely upon the offline-like reconstructed physics 
objects and include  algorithms based upon the 
data in byte-stream format  or  based upon specialized  
raw-data  structures.   
These new algorithms break the factorization of the 
data preparation and the algorithm execution stages.
The introduction of new type of algorithms is reflected in  
an extension of the list  of possible trigger  elements.

All  algorithms are grouped into layers. The main 
goal of such grouping, reflected in the proposed 
data-selection software framework 
\cite{TDAQnotes},
is to create a   
fine-structure of latencies of the event selection process.

The concept of algorithm layers is new.
The eigenstates of each of the algorithm layer are specified in terms of 
{\it the type} 
and {\it the granularity} of data 
they are using,  and in terms their {\it  function} in the data selection.
The data type reflects the stage of the ROD-data unpacking and
preparation. Three
data types are defined in the proposed model:
\begin{enumerate}
\item 
the byte-stream data containing the fixed format and position  LVL2  summary blocks;
\item 
the raw-data objects; 
\item
the reconstructed-data objects.
\end{enumerate}
The allowed granularity of the data which are requested by an  algorithm 
is: 
\begin{enumerate}
\item
a ROD;\footnote{Read-Out-Link (ROL)-unit  would  be  a better, more stable definition
the minimal data unit, independent of possible evolution of 
the TDAQ system;  this is  discussed further in \cite{TDAQnotes}.}
\item
a Region-of-Interest (RoI) group of RODs;
\item
a predefined (constant) group  of RODs;
\item
the full detector.
\end{enumerate} 
The type and granularity of the data used by an algorithm determines
unambiguously how they are aggregated into   
algorithm subsets referred to  in this series of notes as {\bf layers}.
The algorithm  layers are the basic entities (algorithm quanta) which could be 
implemented on any  of the LVL2 processors (LVL2PUs)   or 
Event Filter (EF) processing tasks (PTs).

Each  data  selection algorithm fulfills  one of the two functions:
\begin{enumerate}
\item
Verifying,  if the full (partial)condition corresponding to a trigger element 
is fulfilled (this includes verification of the trigger-menu-predefined {\it veto-elements});
\item
Flagging of   {\it infected data structures } in a way which is independent of the 
preloaded trigger menus. The algorithms 
fulfilling this functions are called in the series of 
presented notes the  {\it T-algorithms}. 
\end{enumerate}

In the proposed model, 
the allowed variability range
of the event selection tools
is confined to a  discrete set of allowed trigger signatures 
and a discrete set of allowed algorithm layers.  
These  eigenstates will be used to define the eigenstates of the  
event-selection-framework configurations discussed in the following section.
The concrete implementation of the model is based upon  the  following seven algorithm layers:
\begin{enumerate}
\item 
The   LVL1 Event Topology Algorithm Layer based upon the Region-of-Interest (RoI) data record 
and upon the trigger system ROD data in the byte-stream form;
\item
The Event Topology Algorithm Layer based upon a group of the ROD data summary blocks;
\item
The Data Validation Algorithm Layer based upon the individual ROD data summary blocks;
\item 
The Electronic Noise and the Beam-Background Algorithm Layer based upon the individual ROD data blocks 
in the byte-stream form;
\item 
The Look-Up Table Algorithm Layer based upon data in the byte-stream form
coming from a group of RODs; 
\item
The Detector-Monitoring Algorithm Layer based upon the raw data objects
derived from the byte-streams of the group of RODs;
\item
The Reconstructed Data Driven Algorithm Layer based  upon the reconstruction input
objects derived from the byte-streams of the group of the RODs. 
\end{enumerate}

The detailed specification of the model of the 
data selection tools  and organization of 
event selection process is presented in a  
dedicated  note \cite{TDAQnotes}.

Within the proposed event selection framework,  each 
LVL1-triggered  event will be  confronted 
with a dedicated selection path composed  
of the  algorithm layers. Such a path is  dynamically configured
on event-by-event basis. An optimal path reflects, 
simultaneously,  the actual data-taking 
conditions and specific wishes of the physics group interested in 
analyzing events of a given type. The freedom in composing 
the path of an event within a  multilayer selection structure in 
a physics-goal dependent way leads to physics oriented
partitioning of the TDAQ system discussed below.

\subsubsection{The eigenstates of the TDAQ-configurations}

The TDAQ system is decomposed into identical  slices.
Such a decomposition is determined, at present \cite{TDR_TDAQ},  solely by 
the  hardware organization of the TDAQ system.
Each of the  LVL2PUs 
aggregated in a  slice
runs the same  standard  event selection software-framework  based upon the 
same run-configuration algorithms and processes  any of the  LVL1-triggered event.
Similarly  identical clone-like organization is foreseen for the 
EF PTs   and the SFIs, where the event building tasks are performed.

In the  model presented in this series of notes,  the clone-like  slices of the TDAQ system are 
dynamically mapped onto  the  {\bf physics-goal optimized  slices}.
The concept of physics oriented TDAQ slices is new.
A physics slice consists  of a subset of the LVL2PUs, 
EF PTs,   and SFIs. This association is virtual, task-driven, 
rather than hardware-driven and can be implemented almost 
effortlessly within the present  hardware architecture.   
Each of the physics slices has its own 
identity determined by the 
type of the LVL1-accepted events which will be 
directed by the LVL2 supervisors 
to this slice.
The  type of event is unambiguously defined by the bit pattern of the
RoI record.
Mapping of the  physics
slice structure  on to the physics group structure 
is highly nontrivial and can be done in several ways. 
A concrete example of  mapping 
 is proposed in a dedicated note \cite{TDAQnotes}. 

The slices may be considered as a physics working group encapsulated
playground,  where their members  will be allowed, within a well defined  rules, 
to make real-data exercises 
and eventually converge to 
their physics-optimal detector-data handling schemes.
Each of the physics slices is given the freedom to implement, if necessary,  its
own event-selection  framework 
using the full set or a fraction of algorithm layers.
Such a  framework could be optimally adapted
to both the concrete physics-goal and to the actual 
data-taking environment. It may include a particular
run-time configuration of standard algorithms confined to one slice.  
Each of the physics slices is also given 
the freedom of choose the implementation place of the chosen subset of the  
algorithm layers between the  LVL2PUs  
and the  EF PTs.
Last,  but not least,  each of 
the physics slices is given the freedom 
to choose 
the event building mode 
(i.e.,  the event data which will be permanently  stored).

The concept of physics slices is central to the proposed model of the 
TDAQ architecture. This concept implements a vision
according to which  
the quest for the physics goal-oriented flexibility of the data-taking 
process will eventually result in the physics-specialized  partitioning of the overall 
TDAQ capacity.  
The effect of such partitioning is that 
the coupling of the data-taking domain to the physics 
analysis domain will inevitably be  enlarged
and thus needs to be precisely defined (modeled).

The physics slice eigenstates of the allowed TDAQ-configurations are expressed   
in terms of: 
\begin{itemize}
\item 
the list of  physics slices - each of slices being unambiguously 
defined by the allowed LVL1 type  of events which could 
be processed on the slice;
\item
the assignment method of the LVL1-accepted events to the 
physics slices; 
\item 
processing capacity 
and event building capacity of each of the the slices;
\item
configuration of the event-selection framework on each of the 
slices;
\item
run-time configuration of algorithms on each of the slices;
\item 
the allowed event-building-modes implemented on the slice.
\end{itemize} 

The full spectrum  of TDAQ configurations,
specified in terms of the above items, 
can not be confined at present to a fixed set of 
eigenstates.
The modeling of the TDAQ configuration variability,  presented in 
a dedicated  note \cite{TDAQnotes}
is restricted,  at present,  to modeling of the 
convergence process to the optimal asymptotic 
set of eigenstates. In the  modeling presented in  Ref. \cite{TDAQnotes}, 
the convergence process is assumed to be driven
by both the dynamic growth of the overall
event selection capacity and of the data collection capacity 
as well as  by the evolution in the research scope.

\subsection{The Granularities}

\subsubsection{The detector granularity} 

The projection of the data scenario eigenstates onto the detector 
granularity consists of  defining the minimal partition 
of the detector,  where  a given scenario is active within a given
time interval.
In the proposed model,  the data content and the 
data compression methods are  allowed to take any of the 
allowed eigenstates on channel-by-channel basis. For example,
the Tile-cal (LAr)  channels with very large energy depositions
may contain the ADC samples. 
The zero suppression eigenstates are allowed to vary  
also on channel-by-channel basis. Since, within 
the present design of RODs, the possibility
of running the local-topology-dependent zero
suppression schemes is excluded, this option   
is introduced mainly keeping in mind a long-time
evolution of the ROD system. At present,  the transitions
between the zero suppression eigenstates can be realized
with the ROD-by-ROD granularity. The 
addressing mode eigenstates  and the ROD summary block eigenstates
are confined to the  on ROD-by-ROD granularity.\footnote{It 
is implicitly assumed that the sub-block structure of the ROD data,
which is  driven 
by the assignment  of channels to a given DSP, 
will respect a ROD-coherent scenario, even if the 
fixed format data summary information will 
be distributed within  
the DSP-based sub-block structure of the ROD data. This is
discussed further in a  dedicated note \cite{TDAQnotes}.}

\subsubsection{The TDAQ granularity }

The minimal unit of the TDAQ granularity in the proposed model is the 
physics-slice. The LVL2 and EF trigger menu eigenstates, the eigenstates
of implemented algorithms layers, the data selection framework
eigenstates, the event building eigenstates,  and the run configuration eigenstates 
of selected algorithms are allowed to vary on slice-by-slice basis.

\subsection{The Dynamics}

\subsubsection{Quasi-static implementation}

In the quasi-static implementation of the model, the  physics  run is the basic time unit.
The transitions between the eigenstates of
the variability domains are allowed  
on the run-by-run basis.
In the quasi-static  implementation,   
the shift crew is given  the responsibility to choose  a particular
set of data-taking eigenstates on the basis of the previous 
run experience and on the basis of anticipated detector performance 
in the  actual data-taking environment. The chosen set  
of run-defined-eigenstates is written to the configuration database 
and drives subsequently the configuration 
dependent features of the data reconstruction process. 
A change in the data-taking environment results in stopping 
and starting the run with the new eigenstates.

The most simple version of the static model
is equivalent to implementation of  
of the present minimal coupling scheme discussed in  Ref. \cite{TDAQnotes}.
This version is based upon a fixed  and uniform-over-the-whole-detector  
data scenario. 
A  single algorithm layer, based
upon the reconstructed objects, is implemented on the TDAQ 
LVL2PUs  and EF PTs.
The software framework of the data selection, 
the run time configuration of algorithms,  and the event building 
mode is frozen during the physics runs and applied to all 
types of LVL1-accepted  events. 

The quasi-static implementation will not be discussed here. 
It is mentioned here to show that the presented 
model contains the currently  implemented data-taking mode.

\subsubsection{Dynamic implementation}

The dynamic implementation of the model discussed  below reflects a compromise between 
the degree of implementation simplicity and  the degree of  flexibility 
of the data-taking process.

It is based upon three 
time-scales: the event-by-event time-scale, the run-by-run
time-scale,  and the period-by-period time-scale.
The modeling consists of defining the minimal  
time-scale at which the 
transitions between the eigenstates 
of the variability domains may  occur
and of specifying the allowed mechanisms activating these
transitions.

In the dynamic implementation,  the 
data scenarios are allowed  to change on event-by-event 
basis. The ROD-granularity transitions
are driven by 
the bit pattern of the
LVL1 trigger word.\footnote{The LVL1 trigger
word is distributed by the TTC system to each of  the RODs of 
the ATLAS detector. The implementation of such a system
was proposed in Ref. \cite{Annecy} and is discussed
in the subsequent note of this series \cite{TDAQnotes}.}
The channel-by-channel granularity transitions are restricted,
given the present capacity of the ROD system, to those driven by 
the channel content.
The eigenstates of the LVL1, LVL2,  and  the EF trigger signatures,  
the eigenstates of the implemented algorithm layers 
and the run-time configuration eigenstates  implemented
on each of the TDAQ slices are allowed
to vary  on run-by-run basis. 
The number of the TDAQ slices 
and the corresponding data-processing 
capacity of each of the slices, the method of assignment of
the LVL1 accepted events to the slices, the configuration 
of the event selection framework and event building 
implemented on the slice are allowed to change on period-by-period
basis. The run- and period-dependent settings of the TDAQ system,
the trigger menus,  and  the implemented layer structure  
define the spectrum of allowed environments in which 
the events are dynamically selected.

In the  scheme  developed in  \cite{TDR_TDAQ}, 
the selection process of each of 
the LVL1-accepted events is fully pre-encoded into the trigger
menus specifying the sequences of algorithms to be run 
and conditions to be met to accept  a particular event.
This process is independent of the event features (secondary
ROIs)  and of the actual data-taking environment. 
For example,  a LAr-coherent-noise event
fulfilling and the 
genuine physics event 
fulfilling the same LVL1 trigger conditions will be treated with the 
same full sequence of algorithms even if the former could be 
recognized very quickly as the noise event.

In the dynamic implementation of the model 
presented here each LVL1-accepted event 
is directed to a predefined TDAQ slice 
and exposed to slice-dependent event-selection framework. 
Each 
LVL1-accepted event will be confronted there 
with a dedicated selection path composed out 
of activated  algorithm layers
and slice-activated HLT trigger menus.  An optimal path reflects, 
simultaneously,  the sensitivity of the rate of the LVL1-accepted events 
to the data-taking environment  conditions,  and specific wishes of the physics group interested in 
analyzing events of a given type.
This functionality can not be provided within the {\it menu-only}
based system,  even if specialized trigger veto signatures
are implemented.

The assignment of  LVL1 accepted events  to a particular
TDAQ slice is driven by the Region-of-Interest-Builder (RoIB)
 record transmitted to the 
LVL2 supervisor. The LVL2 supervisor  balances the load of 
events within the fraction of its processors which are 
associated to one physics slice. The inter-slice
balancing has to be made externally and needs to be dynamically 
established by a laminar, slow and experience driven 
slicing process.  
 
The active algorithm layers implemented on the slice 
determine the latencies of decision steps in rejecting and accepting 
events. 
The activation of the T-algorithms on a particular slice 
is driven by the local, slice-specific  dead time.

\section{The data-taking model and gauge interactions}

\subsection{Gauge analogy}

In the heart of the proposed model is the delegation 
of the adaptive capacities of the ATLAS sub-systems
to specialized variability domains. Within these domains, 
correlated actions have to be taken by each of the 
sub-systems,  both at the time of constructing the data-taking strategy,   
and later, in real-time processing of each of
the LVL1-accepted events.
The efficiency 
of these actions can be quantified in terms of   the 
following four  quality 
criteria:
\begin{itemize}
\item 
The stability of  the data-taking process with 
respect to environmental changes
(the detector operation efficiency);
\item 
The stability of  
the analysis results of each of the physics working groups with 
respect to transitions between the eigenstates of the 
variability domain
(the measurement precision and data selection efficiency);
\item 
The ease of  implementing  of  new
research directions;  
\item  
The readability of the pattern of sub-system couplings.
\end{itemize}
Satisfying 
the latter quality criterion is highly non-trivial in  the {\it multi-component} 
environment of a large experiment.
The approach, which is 
advocated  in the proposed model, follows  the 
adaptation mechanisms of stable biological systems which   
could  be modeled using the concepts of gauge models.
 
\subsection{Gauge and matter fields}

The variable  data-selection, data-collection and data analysis environment 
is represented by  the evolving strength  of the 
three external ``gauge'' fields,  each of them 
varying with its characteristic time-scale: period by period (e.g.,  research goals), 
run-by-run (e.g.,  LHC machine status) 
and event-by-event (e.g.,  electronic noise) time-scale. 
The data collected by the experiment 
depend upon the actual configuration and strength of these fields.

The  collision events are analogous to the ``matter'' fields. 
For  a given data-selection, data-collection and data analysis  environment 
their  `` elementary particle-representations'' 
are  specified by the corresponding set of 
eigenvalues of the variability domains, i.e. by the 
projection of the ``matter fields'' onto the eigenstates
of the variability domains.

The physics results derived from the concrete particle representations
of the matter fields (i.e. from the front-end electronic data registered in RODs for the selected and reconstructed bunch-crossings)
are required  to be invariant with respect to the 
transitions between the eigenstates of the variability domains
(in analogy to freedom 
of the phase rotation of the matter fields in the gauge theory).
This freedom is used, in the proposed model,  to effectively 
absorb  (``rotate out'')  those of external fields which represent
a gauge-dependent (unphysical)  
perturbations of the data 
selection and data collection  process 
upon which the physics results 
are required  not to depend.   

\subsection{Particles and ghosts}

The most optimal set of eigenstates 
of the variability domains
used in construction of the ``data-taking elementary particles''
 is the one 
which allows  for absorbing, in their physics-results-invariant  ``rotations'',
all the unphysical components (``pure gauge configurations'')
of each of the three external fields, in particular 
the unphysical component of the ``strongly interacting field'',
affecting the data-taking process  at the event-by-event time scale and at the  
minimal detector and TDAQ partitions-scale. 
The ``elementary particle'' set,  advocated in this note represents
the first approximation and may very likely  
include an excessive  rotational freedom giving  rise to unnecessary, ``allergic'' reactions
of the data-taking immunological system to ``false alarms'' . The 
inevitable ``ghost-like'' degrees of freedom of
the corresponding gauge fields  
would  become, however,  extinct very quickly
during the adolescence period of the data-taking 
process.   

\subsection{"CKM-like"  rotation}

The  eigenstates of the variability domains 
reacting to the ``strong'', event-by-event-varying   field
are not, in the model presented here,   the eigenstates of the two remaining weaker fields 
affecting the data-taking  at  run-by-run  and period-by-period time-scales.
These latter eigenstates  are constructed as combinations of the strong 
interaction eigenstates.
Such a  rotation of eigenstates, assure decoupling of the ``strong interaction field'' from 
the remaining two  fields.  One of the consequences is that  the symmetry 
governing the event-by-event gauge  invariance of the ROD data  
becomes a  hidden symmetry
(in an  analogy to the SU(3)-color symmetry of the 
Standard Model lagrangian\footnote{
The concrete implementation of this scheme, 
described in detail in a  dedicated note \cite{TDAQnotes}, 
is based on the {\it minimal data-quantum principle},  which assures
an event-invariant presence of the dedicated ROD data structures for  
the online (LVL2 and EF) processing of events.}).    
 
\subsection{Gauge interactions}

In the gauge theory,  interactions of the
matter fields with gauge fields are generated by 
extending the global phase invariance to the 
local ( space-time-point-dependent)  phase invariance.
In an analogous way 
the  interaction between the experiment  sub-components,
represented by the eigenstates of the variability domains, 
are fully specified by 
extending the global rotational invariance 
of the eigenstates of the variability domains (full detector
``space''-scale, run-by-run time scale)  
to the the local invariance   
(partition ``space''-scale  and bunch-crossing time scale).
 Such a  scheme
provides, in its practical implementation,  a very clear,  
effective,  and precisely-defined method of introducing the couplings 
(interactions) between the experiment sub-components
both at the time of constructing the data-taking framework 
and subsequently 
in the  real-time data-taking environment.
At the time  of constructing the framework,
these correlations  
need not to be imposed arbitrarily upon  each of the sub-systems 
but could be confined to static fixing of 
the eigenstates and fixing of invariance rules 
of transitions between the eigenstates on slice-by slice
basis.
At the time of the data-taking, the  implicitly  build-in  correlations will be 
automatically  activated   on event-by-event basis
by  the event-path  or, in a  concrete representation,   by fixing the ``event-gauge''
in terms of the bit pattern 
of the LVL1 trigger word and of  the bit pattern of the RoI record. 

\subsection{Gauge symmetry breaking}

In the proposed model the ``gauge symmetry" is a broken symmetry.
The symmetry breaking pattern is driven mainly by the hardware
capacity of the data selection and the data collection systems. The gauge symmetry 
is restored  in the limit of infinite detector resources.
For example,  while events  collected  using the ROD transparent eigenstate and
using the full event building eigenstate 
(the ``zero mass particles'')
can be propagated quasi freely 
to  any analysis,
events  collected using the reduced-data-content eigenstate
in reduced fraction of sub-detector partitions
 (the ``high mass particles'')
 can be propagated
freely only within a  physics-slice-confined analysis methods.

\subsection{Gauge-independent and gauge-dependent analyzes}

Physicists exposed to  purely offline analysis environment 
and analyzing  their physics-group 
TDAQ and grid slice data could be unaware the data-taking environment 
and the  proposed model altogether. 
In other words,  the gauge-dependent-path 
of (her)his favorite events,  from the LVL1-accept decision 
up to inclusion in the final physics plot
will be  fully encapsulated for (her)him. 
The encapsulation mechanism in exactly the same like the one for  
the QED-events generated 
in the technically most convenient gauge. Any physicists asking 
a valid physics question must not  care which gauge
is used by the event generator.  At worst she (he) may 
be confronted with very ineffective event generation leading to very large  
analysis errors. 

On the other hand,  a  curiosity-driven physicists  will 
be provided with the full gauge-path information for  each of the 
accepted event. This will enable her (him) to search for
the most efficient gauge event-path, to study the gauge 
symmetry breaking phenomena, 
and to  modify  the gauge invariance rules implemented
on his (her) physics group TDAQ and grid slice to 
minimize the measurement systematic uncertainty.

\section{The first ``three minutes'' of an event}

\begin{figure}
\begin{center}
\leavevmode
\epsfig{file=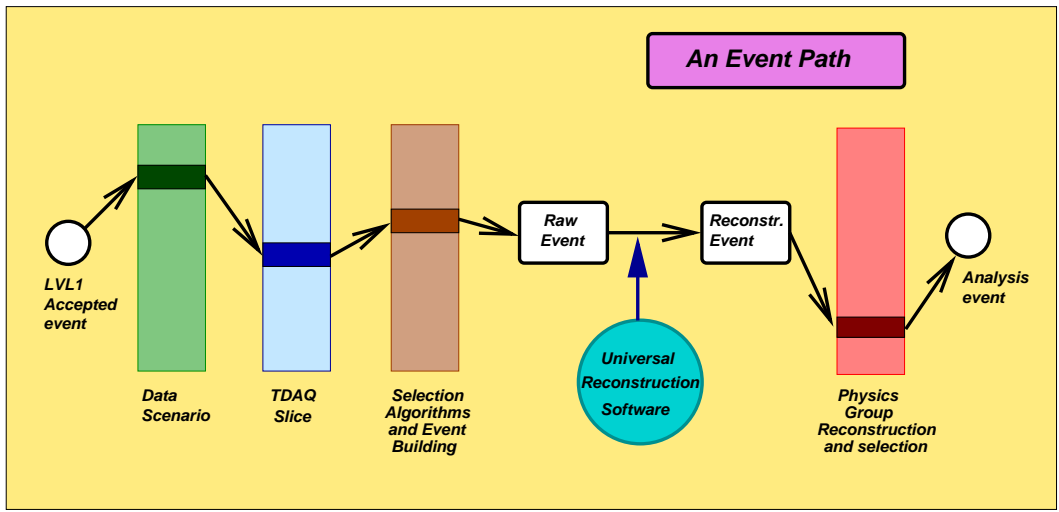,width=14cm}
\vspace{1.0cm}
\end{center}

\caption{An ``gauge-path" of an event.
}
\label{path}
\end{figure}

Fig.  \ref{path}   illustrates the proposed model 
using in   a  practical language. The 
The decision of the LVL1 system to stop the pipelines results 
in transferring the bunch-clocked data to the RODs of each 
of the sub-detectors. The LVL1 trigger word distributed to 
each ROD by the TTC system steers the processing 
of the events by guiding the  choice of 
the content and structure of the ROD data blocks (the  data 
scenario eigenstate). The ROD bunch-crossing-tagged data are formated and sent 
to the corresponding ROBINs over s-links where they will 
wait for the action taken by the LVL2 trigger system.

The action of the LVL2 supervisor is steered  by the 
received  RoIB record. The supervisor delegates to  
a  LVL2PU
the  task of processing the event. This  delegation 
is based upon  the RoIB record. The  
L2PU belonging to a given TDAQ slice
receives only  those of the events which are tagged by a predefined
list of the RoIB bit patterns. The TDAQ slice processors 
can be recognized mainly in terms of the implemented
event selection framework. 
The event is exposed, at the level of the TDAQ-slice processor,  to 
the  slice-dedicated trigger menu and 
the slice-specific data selection algorithms layers. The 
LVL2PU of all but monitoring slice 
use the data-scenario-invariant fraction of the ROD data. The  event 
can be rejected following each layer of algorithms. 

The decision, following which selection
layer the event is built by the slice SFI or rejected,  
is driven by the slice-dependent 
trigger menus. The type of event building 
(full versus  partial, dynamic versus  static) is  implicitly 
encoded in the RoIB record and,  in specific cases,  into 
the LVL2 trigger decision.

The built event is exposed subsequently 
to the  the EF slice-processor algorithm layers. Depending 
upon the slice-dependent  implementation  of algorithm layers
(LVL2 and EF sharing),   the EF processor tasks may or 
may not be guided by the LVL2 results. The accepted EF events
are permanently recorded using the byte-stream form of the data.

The event which passes 
all the above selection stages  is 
subsequently  reconstructed with the Universal 
Reconstruction Software (URS). Its data    
structure is,  irrespectively  from which slice
it is coming from,  invariant - the only difference
is confined to  the data size. 
%\newpage
%width=10cm,bbllx=-29988 ,bblly=274, bburx=1012 ,bbury=30780}
%bbllx=-29988 ,bblly=274, bburx=1012 ,bbury=30780}
The event size is  determined by 
number of the detector
partitions,  which were present in the read-out
system at the time of LVL1 trigger decision and 
which were subsequently chosen by SFI for  the event building.
The reconstruction-input events are depicted as ATLAS raw events. 

The URS, contrary to the HLT reconstruction 
software,  uses the best available data 
using  automatic scenario-dependent decoding procedures.
The URS is limited  to the 
basic reconstruction functions, and stops at the level 
of reconstructed objects  tracks  and clusters.
The reconstructed events are   stored in a way which optimizes
the retrieval speed of a subsample of events 
originating  from a given TDAQ slice.
These events are depicted as the ATLAS  Reconstructed Events.

The physics group selection and the physics-group 
dedicated reconstruction 
of events, in terms of the most appropriate objects for 
a given physics task,  
uses  the Physics Group Software (PGS).
Technically,   any group may use the data coming 
from any slice. The best quality data 
will be those coming from the physics group-dedicated slice configuration..
The PGS records 
n-tuple (root analysis objects ) based upon more   refined selection of events
and containing the final group-specific analysis objects.

\section{Examples of adaptation mechanisms }

\subsection{Anticipated-discovery and generic hot-line events}

Hot-line events are defined as potential discovery events which must be 
streamlined to a dedicated storage and to a dedicated analysis path.
The LVL1 trigger signatures for these events may be confined to 
{\it the anticipated discovery scenarios}  \cite{PhysicTDR}.
They may be also
defined in a more {\it generic} way: for example,   as events containing 
all possible LVL1 signatures,  which are kinematically beyond the reach 
of the FNAL experiments.  
For these   events,  the  most complete 
sub-detector information  (ROD-transparent eigenstate)
for all the available detector partitions 
must be recorded.\footnote{Partially missing electronic information for the 
hot-line events hampered full  understanding of the 
isolated muon events observed by the H1 collaboration
\cite{isolated}. (In this context, note  the merits  of  the event-by event 
transitions in the data content, inherent to the 
proposed  data-taking model, in  avoiding 
the data-flow  bandwidth constrains while  handling the hot-line
events).  }  

The {\it anticipated-discovery}    (hot-line)  events, based upon inclusive 
and double inclusive  high-$E_T$  signatures (small multiplicity of RoIs)
have a residual chance to be noise-infected and/or machine-background-infected. 
Moreover, even if, in the initial phase of the detector operation,
the capacity of the LVL2 and EF farms, and data collection 
system are  reduced (staged),  the selection process of these events
must not  suffer. These events  should be allowed to occupy the 
large-decision-time tail  of the event latency distribution.
 
The  event selection eigenstate, implemented on the 
TDAQ slice where the LVL1-accepted, anticipated-discovery
events are directed ({\it  the ``A-hot-line-slice''})
may thus be based exclusively upon the Reconstruction Data Driven Algorithm Layer
and upon the classical trigger menu  \cite{TDR_TDAQ}.
The chosen eigenstate of the event selection framework  
is  the presently implemented event selection framework.
This framework is characterized  by  
 a nested ROD data access within multiple loop over menu-sequence-table pairs,
sequences, valid TEs and algorithms \cite{Markus}.
The   chosen `A-hot-line-slice''   run-time configuration of the algorithms is  
the one which  is  efficiency rather than purity
driven. The event building eigenstate is the one covering 
all the sub-detector partitions, being in the read-out mode  at the LVL1 decision  
instant  of the run.

A dominant fraction of generic hot-line events can not be identified
on the basis of the inclusive and double inclusive signatures.
This fraction contains those events with  multi-RoI signatures which 
will not satisfy any of the inclusive selection criteria (e.g., an event 
with two $p_t =10~ GeV$ electron RoIs, one $p_t=8~GeV$ muon RoI and three  $p_t=20~ GeV$ jet
RoIs),  and exclusive events (e.g the high transverse energy events  with the energy dispersed
over large number of jet RoIs).
These events can not be selected efficiently by the  presently implemented   event 
selection framework eigenstate  because both  the fraction  of 
the genuine physics events having such 
signatures will be low and unstable (beam-background
and electronic-noise dependent) and because the even accept/reject 
decision will be too long due to combinatorial overhead.
An  attempt to first reconstruct 
the RoIs in terms of particle hypothesis would waste in vain the available
processing power of the farms.  

The selection process of multi-RoI  events must  involve pre-processing  
of the data with the fast data-cleaning algorithms
followed by RoI-by-RoI-based data-verification process. In the initial stage 
of the detector operation,  these data-cleaning algorithms must be implemented 
on the LVL2 farm.  The crucial point here is that the necessary 
change of the selection scheme can not be reduced to introducing 
new signatures for large multiplicity RoI including various  
RoI-dependent veto elements,   but the basic configuration 
of the selection loops must  be reorganized for these events.

The remedy proposed by  the proposed model is the following.
The multi-RoI  generic discovery events  which do not satisfy  the 
inclusive criteria  are directed to the 
dedicated ``G-hot-line slice''. In the LVL2 part of the slice,  the  event selection 
eigenstate is composed out of the three algorithm layers:
the Data Validation Algorithm Layer, the Electronic and Beam Background
Algorithm Layer,  and the Reconstructed Data Driven Algorithms. 

The event-selection  framework eigenstate consists of  a  RoI-by-RoI verification 
process instead of a physics-hypothesis-driven process, as  described in detail in \cite{TDAQnotes}. 
During the noisy periods,  only  a small fraction of events reaches 
the time consuming Reconstructed Data Driven Algorithm layer.

\subsection{Inclusive measurement of high cross section processes} 

In the following example,   the measurement of the jet 
$\eta$ and $E_T$ distributions  down to the lowest limit of the  
kinematical range is considered.
 A standard method 
boils down to  mapping  the $E_T$
spectrum in terms of a set of pre-scaled triggers
and collecting the corresponding events in its data-content-invariant form.
Most of the data collected will never be used and 
the measurement will be  artificially spanned   over large time interval,  thus 
diminishing its systematic precision due to stability 
of the detector calibration.
 
In the present model 
these events are handled in the TDAQ slice dedicated
for large cross section inclusive measurements.
The events occupying the low  $E_T$ part of the spectrum 
are vulnerable to the electronic and beam background
noise. These events are first filtered by the Data Validation Layer and by the 
Electronic Noise and  Beam Background Algorithm Layer. Only when   
they successfully pass each of these  layers  they are exposed to the physics menu driven
Reconstruction Data Driven Algorithm layer. These algorithm layers 
are implemented on the LVL2 ``part'' of the TDAQ  slice.
The event building eigenstate  dynamically chosen  on the
SFI  ``part'' of the TDAQ slice  for inclusive 
events  is the one  collecting  ROD fragments reduced to the RoI associated RODs.
These events need not to be filtered further by the corresponding EF slice
and can be recorded at very high frequency. 

The  above example shows 
how the event-rate-for-event-length trade-off could  work in practice
in the proposed model. 

What if someone wants to study the  particle multiplicity associated
with a jet  of 20 $GeV$ of  $E_T$? (S)he would have to join 
the slice-team, and following the period of optimizing the efficiency 
of inclusive selection of jets,  (s)he could ask for   a transition in  the event building eigenstate 
to get all the info he needs. If (s)he would eventually find out,   that 
what (s)he really needs is not only the multiplicity of particles, but the  multiplicity 
of mini-jest in the forward region, (s)he would need to direct events 
to the EF part of the slice and implement  her(his) mini-jets selection methods 
there.  Her(his) work would be local and her(his) bread-and-butter  activity would not disturb
more fashionable endeavors (e.g. super-symmetry searches).
The main point here is that   there is not much sense in storing a large sample of
large cross section  events 
with the full detector information, just in case 
someone would need it later. Collecting statistically   
sufficient sample would take hours, if such a need would
arise. Moreover, collecting them in optimized beam period  could  
maximize the precision for the well-defined physics measurement.

\subsection{Exclusive events}

As another  example let's consider events with large total  transverse energy.
If the LVL1 total transverse energy signature  classifies  them to 
the rare case of hot-line events one can afford that their 
selection is delegated to the EF part of the  hot-line
TDAQ slice. At smaller $E_t$ such a procedure
brakes down.  The threshold where it brakes down will depend  
upon the actual data-taking environment.
In order to absorb environmental changes due to 
coherent noise and  beam related background 
such events are directed
to the TDAQ slice dedicated for exclusive events 
and exposed at first to 
the two fast-selection  layers:
the   LVL1 Event Topology Algorithm Layer based upon the RoIB record 
and the data coming from the trigger system RODs,  and subsequently to the 
the Event Topology Algorithm layer based upon a group of the ROD data summary blocks.

The algorithms in the first layer analyze the 
topology of secondary RoI, which were not used  
in the LVL1 trigger decision,  
and rejects, as much as it is possible,   the coherent noise 
and the beam wall events. This algorithm layer
is implemented on the LVL2 part of the ``Exclusive slice''.
If the event is retained it  is subsequently built by the slice-SFI.  
The  algorithms of the second layer, implemented
on the EF part of the slice access all the 
calorimeter data but unpack and analyze, at first, only the fixed length 
and format ROD summary blocks to 
verify the consistency of the event
(e.g. by looking at the bit pattern of energy-threshold-crossing cells 
and/or looking  at the timing pattern summaries)
Only if the event passes successfully this stage of the 
selection process,  
the classical EF validation of
the  $E_t$  signature in terms of reconstructed objects
is performed.

\subsection{Baked Alaska events}

The following  example illustrates how new analysis ideas
could  be incorporated and how natural
and clash-less  the diversification 
of the physics goals within the proposed data 
taking model could be. 

Let's imagine that one of the forgotten but very attractive Bjorken's
ideas becomes suddenly attractive  to a group of physicists
who would decide to go ``fishing'' 
as intermezzo  of  exhaustive ``gold(higgs)- mining''.
For example  they may embark on searches of 
the ``baked Alaska'' events \cite{Bj}, originating   from a 
disoriented  chiral condensate. 
Such events are expected to be produced in soft collisions 
and can not be selected on the basis of the high $E_T$
signatures. 
Confronted with  the standard  data-taking procedure 
``the group of fishermen's'' will, most likely,  try to analyze a large sample of 
random bunch crossing events with little hope of 
finding anything interesting. Their attempt to 
have their dedicated-event-selection 
scheme implemented concurrently with other 
schemes  will  most likely fail,
because of the large latency dispersion 
to accept/retain their events and the corresponding necessity to change
the global rules of aborting looping processors - indispensable
for other `golden'', according to fashion dictators,  type of events.

In the proposed model,  the  transition to  
an eigenstate of the TDAQ configuration 
consisting of adding a residual-size ``fishing-slice''
would do the work. This slice could  be optimized  
for large latency  selection of interesting 
 minimum bias events. The creation of such a slice could 
allow them to optimize their selection methods in ``non-invasive'' 
way and consequently drastically increase their chances
of seeing exotic chiral states -  almost in the unnoticeable 
way for the mainstream activities.

\subsection{Detector monitoring events}

In the proposed
model,  the detector monitoring events, tagged  by their monitoring 
LVL1 trigger type,  are directed to a specialized  TDAQ slice
on which a dedicated event selection and event analysis framework
is implemented. This slice has increased network capacity  allowing 
for congestion-less  access to  SFI-built  events from external sites. 
Events tagged as monitoring ones are recorded 
using  the ROD-transparent data eigenstate.
They will be recorded concurrently with the physics events.
Since their rate will be  small  full 
n-samples history of ADC counts is  sent from RODs to ROBins.

The monitoring events are classified in the 
model either as bunch crossing clocked events  
or as random events. The former are numbered according to the bunch 
numbering scheme and contain the subsamples of, 
minimum bias events
(numbered according to position of the colliding 
bunches within the bunch-train), pilot-bunch events, and empty bunch events. 
 
The monitoring events are subdivided into the three groups
which will be processed differently. The association 
is random but respects  the optimal relative sizes of each group.
 
For the  monitoring events belonging to the first group,   the LVL2 part of the
slice is transparent. A sub-fraction of these  events is  transfered directly to the 
slice SFIs. These events  are subsequently analyzed locally and/or externally
by the sub-detector experts. The remaining events of the first group  
may be exposed to all or to  a subset of the presently employed 
data selection algorithms. In this case  the selection decision 
is recorded but it is not active. These events help 
in debugging  and monitoring the performance of  the implemented selection algorithms  

The second group of monitoring events is  exposed to the Detector Monitoring 
Algorithm Layer based upon the raw data objects. These raw
data objects are stamped by the  on-line cell identifiers facilitating
their preselection for dedicated monitoring 
tasks of the sub-detector electronics (e.g using enriched sample of particular
noise pattern events). 

The third group  of monitoring events 
is exposed to a special set of monitoring algorithms  included in  
the Reconstructed Data Driven Algorithm Layer and
driven by dedicated trigger menu for monitoring events.
Events selected by these algorithms contain 
particular physics-features, reconstructed using 
a subset of  sub-detectors,  which is used  to monitor the 
performance of the remaining sub-detectors. 
For example,  preselected random events with a muon track 
reconstructed both in the central tracker and 
in the muon chamber provide an unbiased, optimal sample 
 for dedicated studies  
of the response of ATLAS calorimeters to vertex pointing muons.

\section{Conclusions}
 
This note introduces  a novel  paradigm for  the data selection, acquisition and analysis
for a  multipurpose experiment at the LHC collider. While the presently implemented
data selection and data analysis model tries to assure  the most efficient scrutinizing  of the precisely predefined
discovery scenarios for the LHC collider, the  model presented here tries to optimize the data selection 
and the data analysis framework for an open physics goal,  generic research program giving 
a substantial freedom and democracy to the physics groups to impose their exclusive data 
selection, acquisition and analysis methods in a clash-less manner.   
If the diversification of the LHC research program 
will turn out to be necessary and if it will not be confined to expanding (i) the trigger menus and (ii)  the methods of the off-line analysis of the standard data structures, then  the proposed model could be
a solution for the advanced phase of the LHC experimental program.
 
%\section*{Acknowledgment}

\end{document}